\documentclass[11pt,twoside]{article}
\usepackage{./asp2021}

\aspSuppressVolSlug
\resetcounters

\begin{document}

\title{Protoplanet Express, a video game based on numerical simulations}
\author{Jorge Cuadra,$^{1,2}$ Miguel Vergara,$^3$ Bruno Esc\'arate,$^1$ Josu\'e Sandoval,$^3$ and Benjam\'\i n Cordero$^1$\\
\affil{$^1$Universidad Adolfo Ib\'a\~nez, Vi\~na del Mar, Chile; \email{jorge.cuadra@uai.cl}}
\affil{$^2$N\'ucleo Milenio de Formaci\'on Planetaria (NPF), Chile; }
\affil{$^3$Universidad T\'ecnica Federico Santa Mar\'\i a, Valpara\'\i so, Chile.}}

\paperauthor{Jorge Cuadra}{jorge.cuadra@uai.cl}{0000-0003-1965-3346}{Universidad Adolfo Ib\'a\~nez}{Facultad de Artes Liberales}{Vi\~na del Mar}{Valpara\'iso}{2520000}{Chile}
\paperauthor{Miguel Vergara}{miguel.vergarap@sansano.usm.cl}{}{Universidad T\'ecnica Federico Santa Mar\'\i a}{Departamento de Inform\'atica}{Vi\~na del Mar}{Valpara\'\i so}{2340000}{Chile}
\paperauthor{Bruno Esc\'arate}{bescarate@alumnos.uai.cl}{}{Universidad Adolfo Ib\'a\~nez}{Facultad de Ingenier\'\i a y Ciencias}{Vi\~na del Mar}{Valpara\'iso}{2520000}{Chile}
\paperauthor{Josu\'e Sandoval}{josue.sandoval@sansano.usm.cl}{}{Universidad T\'ecnica Federico Santa Mar\'\i a}{Departamento de Electr\'onica}{Vi\~na del Mar}{Valpara\'\i so}{2340000}{Chile}
\paperauthor{Benjam\'\i n Cordero}{becordero@alumnos.uai.cl}{0009-0003-0149-3871}{Universidad Adolfo Ib\'a\~nez}{Facultad de Ingenier\'\i a y Ciencias}{Vi\~na del Mar}{Valpara\'iso}{2520000}{Chile}

\begin{abstract}
Astronomical images can be fascinating to the general public, but the interaction is typically limited to contemplation.
Numerical simulations of astronomical systems do permit a closer interaction, but are generally unknown outside the research community.  
We are developing "Protoplanet Express", a video game based on hydrodynamical simulations of protoplanetary discs.
In the game, the player visits several discs, finds its relevant features and learns about them.
Here we present the current version of the game, discuss its reception, and consider its further development. 
\end{abstract}

\section{Introduction}

Planet formation takes place in proto-planetary discs.  
These are relatively thin structures of gas and dust orbiting around young stars.
During the last decade new observational facilities have revealed the discs' morphologies.
Rather than being uniform, we see that most of the brightest discs around young stars are very complex objects.
They have large inner cavities, spiral arms, rings, gaps, and other asymmetries.

To understand the disc morphologies, astrophysicists develop numerical models, in which planets or stellar companions are sometimes included.
For some discs we know that these objects are there, while in many cases the presence of companions is proposed in order to match the disc features (HD 169142 is shown as an example in Fig.~\ref{HD169142}).
For instance, the gravitational effect of a close binary star will evacuate the inner part of the disc, creating a large cavity.  
It will also excite spirals arms that propagate outwards from the inner edge of the truncated disc.
A Jupiter-mass planet can produce the same kind of effects, but of smaller magnitude, for instance, rather than a large cavity a planet empties an anular gap. 
Both the gas and dust distributions can be obtained from the numerical models.
The gas typically shows spirals better and has a more extended distribution, while the dust produces sharper features like rings and clumps.
The simulation output can be post-processed in order to generate synthetic images, which are compared to the observed data in order to confirm or refine the proposed model. 

\articlefiguretwo{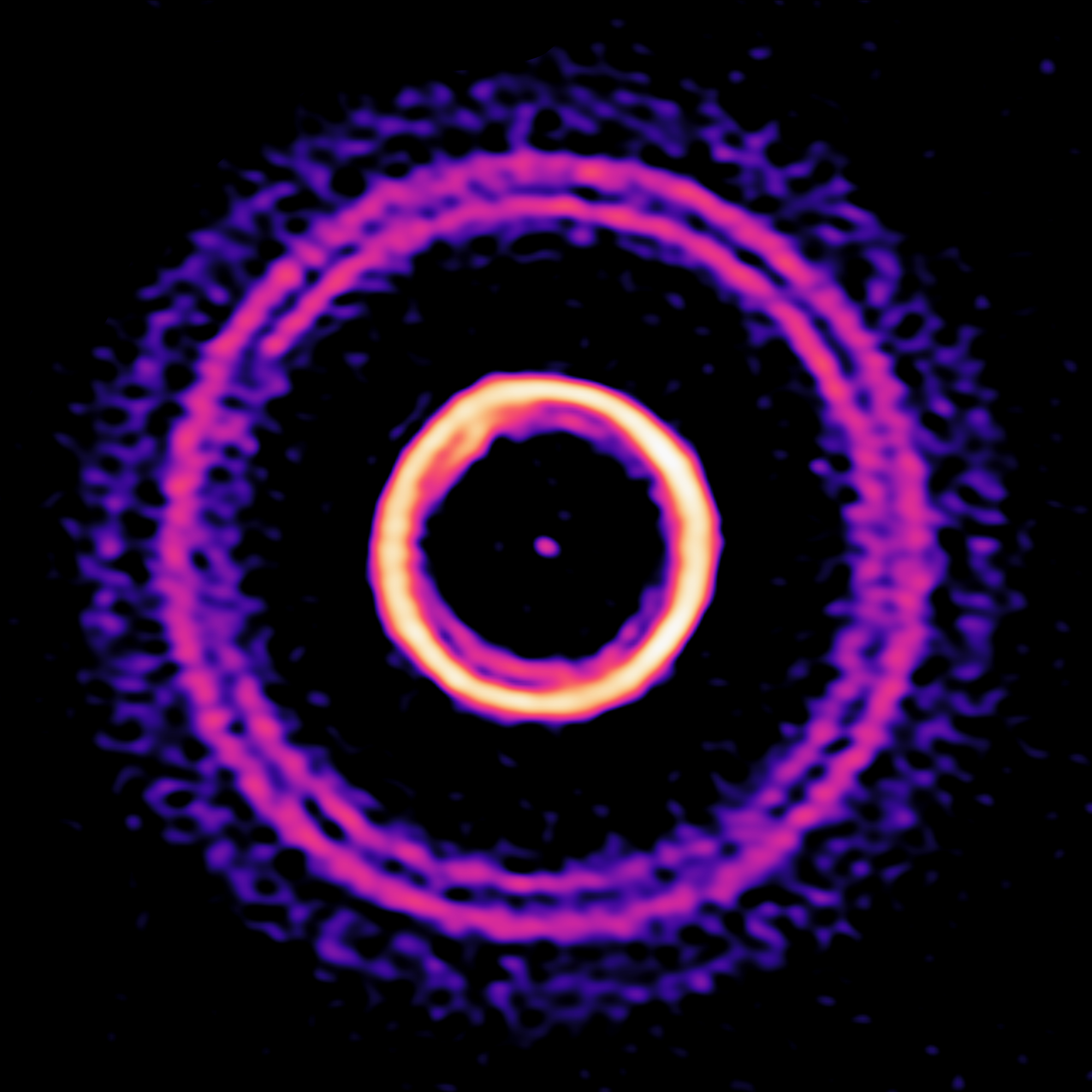}{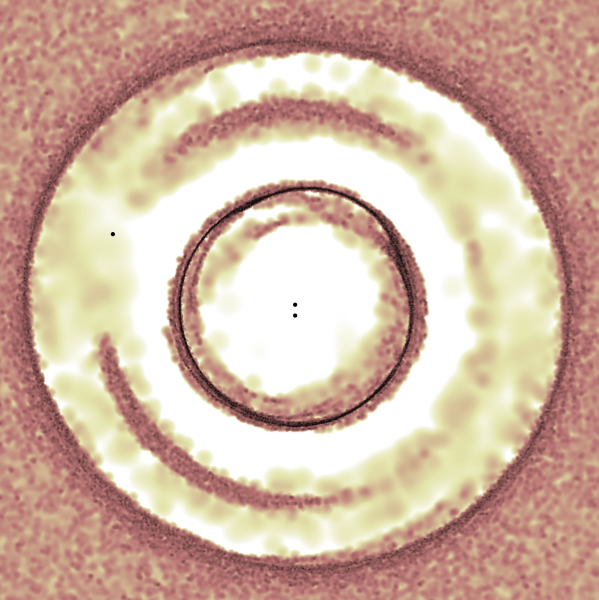}{HD169142}{Dust in the protoplanetary disc HD~169142.  \emph{Left:} ALMA image of the system,  obtained by \cite{Perez19}.  Credit: N.~Lira - ALMA (ESO/NAOJ/NRAO); S.~P\'erez - USACH/UChile.  \emph{Right:} Dust distribution from the simulation of this system by \cite{Poblete22}, which we use in the game (Fig.~\ref{HD169142_game}).}

Observed images of protoplanetary discs are spectacular, and get regularly featured in mainstream media.
On the other hand, the simulation approach to their research is rarely known by the general public.
Additionally, simulations allow for a more immersive exploration of a system, as a three-dimensional domain that can be not only oriented as the user wishes, but also transversed through.
In contrast, astronomical images are typically two-dimensional and do not afford much interaction.
Here we present "Protoplanet express", a video game we are developing to fulfil the possibility of exploring protoplanetary disc simulations.
We hope that the players will enjoy the ride while learning about planet formation.

\section{Video game concept and development}

We wanted to translate numerical models to a format that can be enjoyed interactively by general public.
Our group had experience developing 360$^\circ$ and virtual reality visualisations of Galactic centre simulations  \citep{Russell17, Russell21}. 
Even though exciting for researchers, these visualisations were hard to interpret for the public, as the system was too complex, and there was no associated narrative.
Therefore, we decided to switch to the relatively simpler protoplanetary discs, and to develop a product that feels like a regular video game, rather than a research tool or an encyclopaedic app. 

We also had the aims of staying as close to the actual research as possible, and to maximise the potential of the game as an outreach tool.
Therefore, we decided to use data from recent, peer-reviewed simulations, and also to consider only models of specific protoplanetary discs. 
The latter allows the player to identify the systems and look for additional information, and even to look for them in the night sky. 
We used output data from smoothed particle hydrodynamical (SPH), dust and gas simulations.
SPH is a particle-based technique, so it is straightforward to export the disc structure to be processed further. 
Also, having both gas and dust components allows the player to get a more complete picture of each system.
As an example, Fig.~\ref{HD169142_game} shows how the system HD 169142 appears rendered in the game.  
Notice that while Fig.~\ref{HD169142} shows the dust distribution, in Fig.~\ref{HD169142_game} both the dust and gas are visible.

\articlefigure{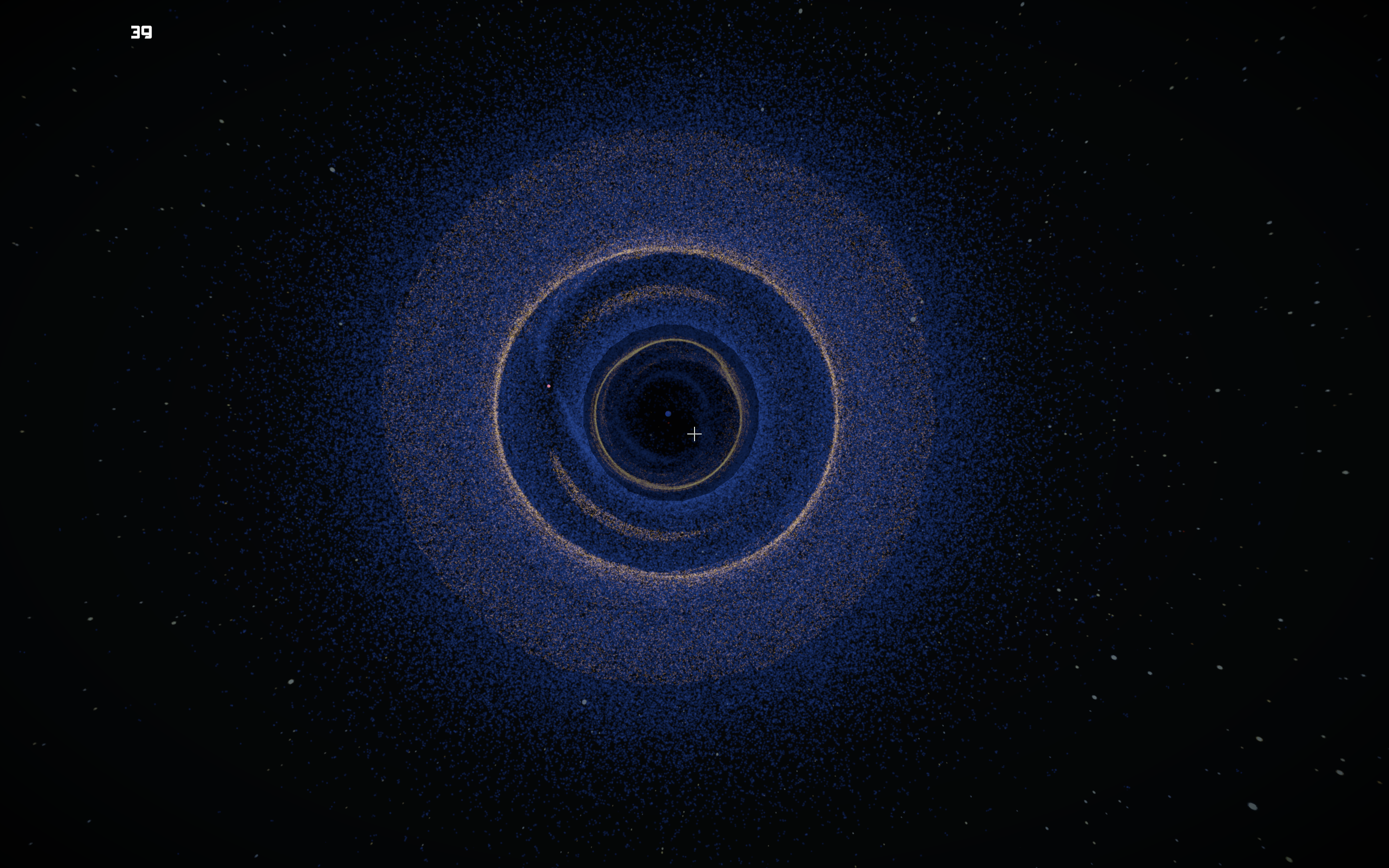}{HD169142_game}{HD 169142 as rendered in the game. Blue shows the gas particles, while dust particles appear in orange (same as Fig.~\ref{HD169142}, right).  A central binary, composed of a blue and a smaller red star, and a pink planet on the left are also visible.}

We used standard tools of the video game industry to design the game.
Blender was used in the modelling of the spaceship and the structure of the simulated dust particles, while Unity was the game engine used to create the game, using its Visual Effect Graph to render the simulated gas particles.
More details about the whole process are deferred to a follow-up publication.

\section{Gameplay}

The game puts the player inside a spaceship in a first-person view.  
The ship features a cockpit, an observation deck, and a chart room.  
From the cockpit the player can control the ship, as explained below.
The observation deck allows a clear view of the protoplanetary disc being visited, as shown in Fig.~\ref{HD169142_game}.
Finally, the chart room has screens with text and images of the disc, plus a hologram showing its 3D structure (see Fig.~\ref{chartroom}).

\articlefigure{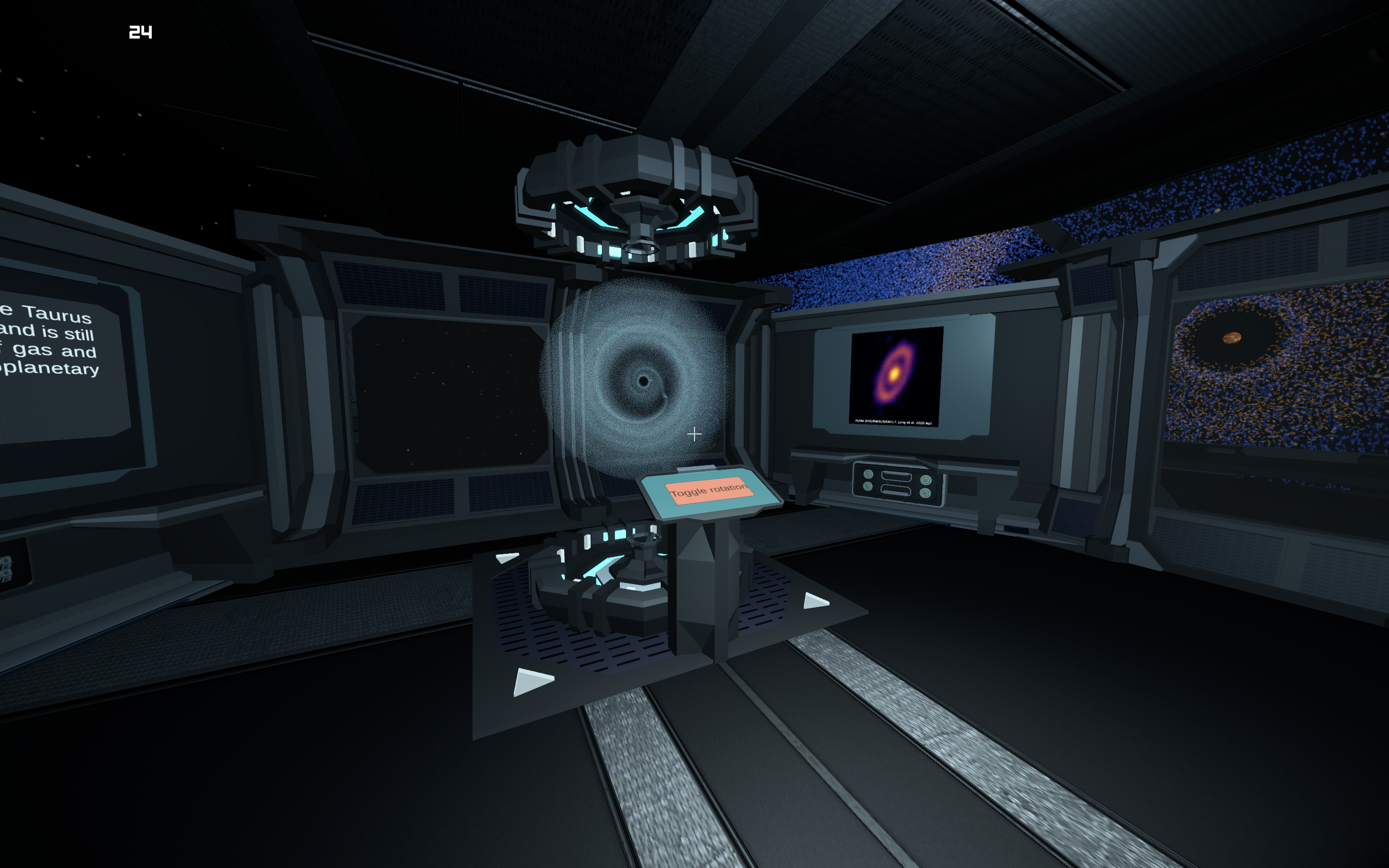}{chartroom}{View of the chartroom.  It features, from left to right, a screen with informative text, a hologram of the disc being visited (DS Tau), and a picture of the disc as observed from Earth.  Through the ship windows the rendered disc can be seen outside.}

When the player is controlling the ship, this can be moved and rotated in any direction.
The player is supposed to explore each protoplanetary disc, finding interesting features.
These can be parts of the disc itself, such as a spiral arm, or the object causing it, as a planet.
When finding each of these features, a brief explanation will appear on the screen, presented by a character based on the main developer of the model (see Fig.~\ref{gameplay}).
After the player finds a required set of features, the level is considered as completed, and they will be allowed to travel to the next system.    

\articlefigure{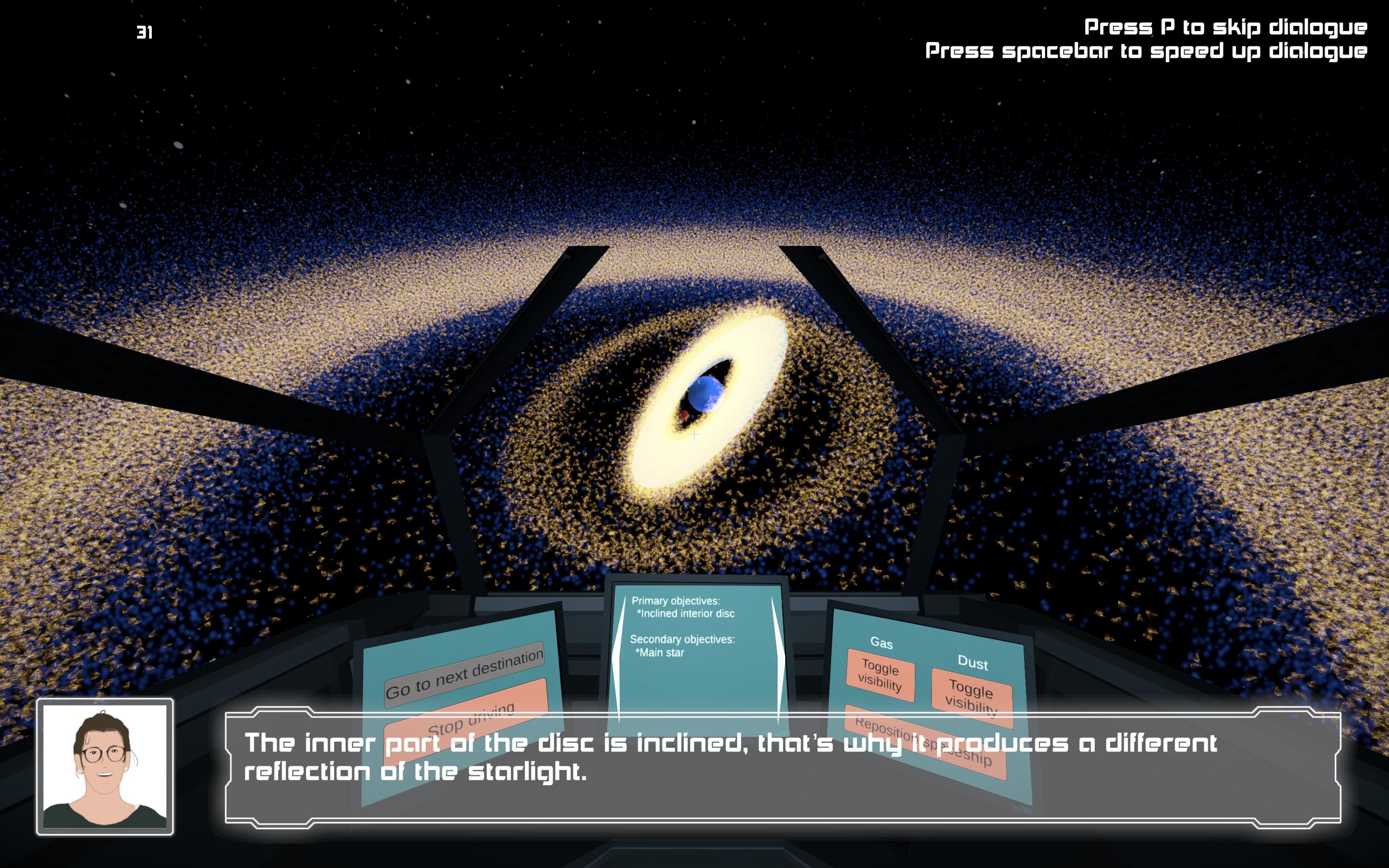}{gameplay}{One of the protoplanetary discs as seen from the cockpit.  When the player finds one of the selected features, in this case the inclined inner disc around HD 143006, an informative message appears.}

\subsection{Protoplanetary discs included}

The current version of the game includes six levels, each corresponding to one of the discs described below.  The citations correspond to the papers presenting the models we rendered for the game.

\paragraph{DS Tau}
In this disc we can observe a low-density gap, surrounded by a denser ring. This gap and ring can be due to giant planet, of approximately 3.5 times the mass of Jupiter, growing in the disc \citep[shown in Fig.~\ref{chartroom} here]{Veronesi20}.

\paragraph{PDS 70}
This system has two actually observed planets.  The disc around the star is truncated in its inner part, likely due to the planets' influence. Finally, we can see a gas influx from the disc to the outermost planet, and a very small disc around it, where moons could be formed \citep{Toci20}.  

\paragraph{HR 8799}
This is an older star, so its gas disc already dissipated. Only a debris disc remains, with the leftovers of the planet formation process. This star also hosts four giant planets, which we observe orbiting around it.   The disc extension is determined by the planet orbits, which remove material that gets close to them.  For this system we did not have a numerical simulation, so we use particle data generated by V.\ Faramaz based on her analytical model \citep{Faramaz21}.

\paragraph{IRAS 04158+2805}
A binary system, with each star surrounded by a disc of gas and dust, all of which is located within the central cavity of a larger, circumbinary disc.
 The gravity of the binary forms the cavity, and also pulls material from the disc inner edge, so it piles up on one side \citep{Ragusa21}.

\paragraph{HD 143006}
This disc has a cavity in its centre. Moreover, the disc seems "broken", with its inner and outer parts having different inclinations. This strange setup can be explained by the gravitational influence of a low-mass star, which creates the cavity and aligns the inner disc, and a planet, which carves the gap that disconnects both parts of the disc, getting them misaligned  \citep[shown in Fig.~\ref{gameplay} here]{Ballabio21}.

\paragraph{HD 169142}
This disc also has a large cavity in its centre. The gas in the disc forms spiral arms, while the dust has a complex structure with rings and gaps. Moreover, the inner ring appears clumpy rather than uniform. As in the previous case, all this complexity can be explained by the gravitational influence of a low-mass star, which creates the cavity, and a planet, which opens up a gap. Both together produce the spiral arms  \citep[shown in Fig.~\ref{HD169142_game} here]{Poblete22}.

\section{Current status and outlook}

We have presented the game at several science fairs, where some of us together with astronomy students have guided the public while trying the game.
We also made a beta version of the game available online\footnote{Currently available at \url{https://questionablegames.itch.io/protoplanet-express}}.  
That version had a questionnaire attached to it, in order to assess the players' experience, but unfortunately we received too few answers to be able to analyse them systematically.
In any case, the feedback we received both in person and online has been useful to make several adjustments to the game.
In particular, we made it more straightforward for the software to register that a feature has been found.  
Originally we required the player to "touch" the feature with the ship, while in the current version the player scans the sky using the cursor.

There are further changes suggested by the players' feedback, and we are in the process of implementing them.
For instance, more direct instructions are needed so the player knows what they are supposed to do, both in terms of what the game is about and which buttons to use.
We are also considering to have two different versions of the game, one for people to download and try at home, which should be somewhat challenging in order to keep the player engaged, and another one we can use at exhibitions, where we want the player to quickly find the objectives and not become frustrated.
At the science fairs we have also realised that the most enthusiastic players are children under 10 years old, so  we are considering to modify the texts and graphics aiming for that audience. 
Finally, our game is designed to run in computers, due to the performance requirements of showing data of roughly a million particles.
However, there is a clear demand for a lighter version that can run on tablets or smartphones.  
We hope to achieve a balance between versatility and faithful reproduction of the numerical models.

Our own short-term plans include having an official, professional-looking release of the game in Steam and other popular platforms. 
After that is finished, we plan to develop a virtual reality version, in which the player can really get immersed in our simulated systems.
In general, we hope developing the video game becomes part of our research routine, so each time we publish a new numerical model, we release a new version of the game in which the model is included.
In that way we hope to keep the general public engaged in this fascinating aspect of astrophysical research.

\acknowledgements Simulation data for the game kindly provided by B.\ Veronesi, C.\ Toci, V.\ Faramaz, E.\ Ragusa, G.\ Ballabio, and P.\ Poblete.
Thanks also to N.\ Cuello, R.\ Nealon, A.\ Dunhill, C.\ Hall, C.\ Russell, and O.\ Guilera.
Figure~\ref{HD169142}, right panel, was created with Splash \citep{Price07}.
Project partially funded by ANID (FONDECYT 1211429, and Millennium Science Initiative Program - NCN19\_171).



\end{document}